# Voltage-induced defect mode interaction in a one-dimensional photonic crystal with a twisted-nematic defect layer

Ivan V. Timofeev,[1,2,*] Yu-Ting Lin,[3] Vladimir A. Gunyakov,[1,2] Sergey A. Myslivets,[1] Vasily G. Arkhipkin,[1,2] Stepan Ya. Vetrov,[2] Wei Lee,[4,5] and Victor Ya. Zyryanov[1,6]

[1]*L.V. Kirensky Institute of Physics, Krasnoyarsk Scientific Center, Siberian Branch of the Russian Academy of Sciences, Krasnoyarsk 660036, Russia*
[2]*Siberian Federal University, Krasnoyarsk 660041, Russia*
[3]*Master Program in Nanotechnology, Chung Yuan Christian University, Chung-Li 32023, Taiwan*
[4]*Department of Physics, Chung Yuan Christian University, Chung-Li 32023, Taiwan*
[5]*Center for Nanotechnology, Chung Yuan Christian University, Chung-Li 32023, Taiwan*
[6]*Siberian State Aerospace University, Krasnoyarsk 660014, Russia*

Defect modes are investigated in a band gap of an electrically tunable one-dimensional photonic crystal infiltrated with a twisted-nematic liquid crystal (1D PC/TN). Their frequency shift and interference under applied voltage are studied both experimentally and theoretically. We deal with the case where the defect layer thickness is much larger than the wavelength (Mauguin condition). It is shown theoretically that the defect modes could have a complex structure with the elliptic polarization. Two series of polarized modes interact with each other and exhibit an avoided crossing phenomenon in the case of opposite parity.

*Corresponding author: tiv@iph.krasn.ru

## I. INTRODUCTION

Photonic crystals (PCs) are optical materials that allow a rich optical response to be obtained by means of spatial inhomogeneity [1]. Defects in such structures with periodicity cause extreme characteristics in optical response. By introducing a tunable defect substance, the optical spectrum can be further controlled. Liquid crystals (LCs) are well-known, promising materials featuring in sensitive anisotropy, dispersion, nonlinearity and susceptibility to external factors, such as temperature, electric and magnetic fields, and light [2–8]. Combining the above concepts, PCs infiltrated with nematic LC as a defect (PC/NLC) were suggested [9–12]. PC/NLC allows manipulation with the defect transmittance peaks to be carried out in the photonic band gap (PBG). Self-organized PC with various types of pitch defects were studied in cholesteric LC as well [13]. Furthermore, an anisotropic Fabry–Perot resonator cavity has two orthogonally polarized mode series [14]. The concept of two mode series (O- and E-modes) was applied to the PC/NLC system to reveal that the mode interaction depends on the mode parity [15-18].

Investigation of a twisted-nematic defect in PC (PC/TN) was initiated in [19]. The average molecular orientation (director) of the rod-like NLC is twisted by 90° using orthogonal planar boundary conditions. This twist provides coupling between O- and E-modes, which leads to elliptic polarization of the defect mode. Our goal in this paper is to clarify the underlying processes behind the experimental data reported in Ref. 19.

## II. NUMERICAL ANALYSES

Numerical routines were used to find both the LC orientation and PC/TN optical response. The first step is to calculate the reorientation of the LC director inside a twisted-nematic cell (TNC) (Fig4. 1(a)) under an applied voltage. The voltage is applied across the cell thickness (along the $z$-axis, see Fig. 2). Before we start our calculation, it is instructive to predict the result qualitatively [20]. Let us roughly assume a LC cell to consist of two boundary layers and a middle layer. Voltages higher than the Fréedericksz threshold [21] force the LC director to lie along the $z$-axis (known as homeotropic orientation), whereas the boundary LC layers remain in planar alignment due to the surface condition. The middle layer then acquires homeotropic orientation and loses azimuthal rigidity. In this case, the azimuthal coupling between the left and the right boundaries vanishes and both boundary layers return to their respective rubbed planes (Fig. 1(b)). The homeotropic middle layer acts as a lubricator. The applied voltage gives rise to the homeotropic lubrication effect. The azimuthal angle $\varphi$ of the LC director varies slowly near the boundaries yet exhibits a steeper behavior in the middle. Increasing voltages lead to the growth of the middle homeotropic layer and the simultaneous reduction of the planar boundary layers (Fig. 1(c)) until the thickness of the boundary layers approaches the electric coherence length $\xi$, which becomes much smaller than the light wavelength.

Minimization (variation) of free enthalpy is a classical method to determine the LC director orientation. While mechanical increment of free enthalpy is negligibly small, it is common to minimize the free energy [22,23]. An analytical solution for twisted LC in the electric field is rather involved [24]. A numerical solution of the boundary value problem using the shooting method [25] is appropriate for in-plane LC orientation, whereas for twisted LC it is sufficient to use a simple general method of gradient descent to the free energy minimum [26]. The elasticity energy of a homogeneous LC layer is expressed as [23]

$$2F_k = \overline{k}_b \tilde{\theta}^2 + \overline{k}_t \tilde{\varphi}^2 \cos^2\theta \qquad (1)$$

Here $\theta$ and $\varphi$ are the angle of the LC director deviation from the substrate plane and the azimuthal angle of the director $\vec{n}$, whose components are $n_x = \cos\theta\cos\varphi$, $n_y = \cos\theta\sin\varphi$, $n_z = \sin\theta$. The tildes in Eq. (1) mean a partial derivative with respect to the cell depth, for example, $\tilde{\theta} = \partial\theta/\partial z = \partial_z \theta$, where $\partial_z$ is the partial derivative operator with respect to the $z$-axis. Effective elasticity coefficients $\overline{k}_b$ and $\overline{k}_t$ are given by

$$\overline{k}_b = k_{11}\cos^2\theta + k_{33}\sin^2\theta, \; \overline{k}_t = k_{11}\cos^2\theta + k_{22}\sin^2\theta, \qquad (2)$$

where $k_{11}$, $k_{22}$ and $k_{33}$ are the splay, twist and bend elasticity coefficients, respectively.

The continuum theory of NLC treats free energy density as a continuous function of coordinate ($z$-axis). A numerical approach converts this function into discrete array of numbers, corresponding to thin semi-homogeneous sublayers. In this study we used 100 sublayers for fairly accurate solution. Numerical convergence was controlled by artificial viscosity. The free energy variation is

$$\delta F_k = \delta_\theta F_k \, \delta\theta + \delta_\phi F_k \, \delta\phi. \tag{3}$$

Here $\delta_\theta = \delta/\delta\theta$ and $\delta_\phi = \delta/\delta\phi$ are the functional derivative operators. $\delta_\theta F_k$ and $\delta_\phi F_k$ are the functional derivatives, known as Frechet derivatives.

$$2\delta_\theta F_k = \left(\delta_\theta \overline{k}_b\right) \tilde{\theta}^2 + \overline{k}_b \left(\delta_\theta \tilde{\theta}^2\right) + \left(\delta_\theta \overline{k}_t\right) \tilde{\phi}^2 \cos^2\theta + \overline{k}_t \, \tilde{\phi}^2 \left(\delta_\theta \cos^2\theta\right), \tag{4}$$

$$2\delta_\phi F_k = \overline{k}_t \left(\delta_\phi \tilde{\phi}^2\right) \cos^2\theta. \tag{5}$$

The numerical approach converts the functional derivative into one-sublayer free energy gradient. Sublayer $z_i$ with angles $\theta_i$ and $\varphi_i$ gives

$$2\delta_{\theta_i} F_k = \left(\partial_{\theta_i} \overline{k}_b\right) \tilde{\theta}^2 + \overline{k}_b \left(\partial_{\theta_i} \tilde{\theta}^2\right) + \left(\partial_{\theta_i} \overline{k}_t\right) \tilde{\phi}^2 \cos^2\theta + \overline{k}_t \, \tilde{\phi}^2 \left(\partial_{\theta_i} \cos^2\theta\right), \tag{6}$$

$$2\delta_{\phi_i} F_k = \overline{k}_t \left(\partial_{\phi_i} \tilde{\phi}^2\right) \cos^2\theta. \tag{7}$$

The five gradient components can be interpreted as elastic forces of an opposite sign. The electrostatic energy of the LC sublayer is expressed as [22]

$$2F_e = -\vec{D}\vec{E} = \frac{-D_z^2}{\left(\varepsilon_\perp \cos^2\theta + \varepsilon_\| \sin^2\theta\right)}. \tag{8}$$

This one-dimensional problem restricts the electric induction to depend on $z$-axis only. Moreover, this dependence vanishes as long as the field divergence is zero.

$$\left|\vec{D}\right| = D_z = const(z). \tag{9}$$

The total electrostatic energy of a cell has to be expressed through voltage $U$:

$$2\Phi = 2\int_0^L F_e = \frac{-U^2}{\int_0^L \left(\varepsilon_\perp \cos^2\theta - \varepsilon_\| \sin^2\theta\right)^{-1} dz}. \tag{10}$$

Again, the electrostatic polarization force of the i-th sublayer is equal to the partial derivative of energy with respect to the angle $\theta_i$ at a constant voltage $U$ on the electrodes. Here we ignore polarization of the surfactant layers. Experimentally, the surfactant cannot be moved outside the electrodes.

Another routine was used for simulation of PC/TN optical response. The case of normal light incidence is considered. To verify the experimental results, we calculated transmission spectra under increasing voltage. The optical response was found using the Berreman method– the transfer-matrix method generalized for an anisotropic medium [27,28]. In the case of isotropic layered media, electromagnetic radiation can be divided into two independent (uncoupled) modes. These are two modes with orthogonal electric field vectors. Since they are uncoupled, the matrix method involves manipulation of $2 \times 2$ matrices to describe propagation of the forward and backward Rayleigh waves. In the case of birefringent-layered media, the electromagnetic radiation consists of four partial waves. Mode coupling takes place at the interface where an incident plane wave produces waves with different polarization states due to anisotropy of the layers. As a result, $4 \times 4$ matrices are required in the matrix method [29,30]. Both orientation and optical routines were implemented in MatLab.

## III. EXPERIMENTAL SETUP AND MEASUREMENT

A layout of the electrooptical 1D PC/TN cell used in our experiments is shown in Fig. 2. It contains two identical dielectric mirrors and a twisted-nematic LC as a defect layer. The multilayer film of each mirror comprised 6 layers of high-index substance, zirconium dioxide (ZrO$_2$) with the refractive index $n_1 = 2.04$ and thickness $l_1 = 0.052$ $\mu$m, and 5 layers of low-index dielectric, silicon dioxide (SiO2) with the refractive index $n_2 = 1.45$ and thickness $l_2 = 0.102$ $\mu$m. These layers were deposited in an alternating sequence onto fused quartz substrates provided with a conductive coating indium tin oxide (ITO). We chose the well-known LC, 4-*n*-pentil-4'-cyanobiphenyl (5CB) with nematic phase between 22.5 and 34°C as the defect material. Planar alignment was obtained by a thin polyvinylalcohol (PVA) surfactant layer. A twist of 90° was achieved by silk rubbing procedure on the alignment layers. Thickness of LC layer is about 14 $\mu$m, the anisotropic refractive indices of 5CB are $n_\parallel = 1.719$ and $n_\perp = 1.536$ ($T = 23$°C, $\lambda = 589$ nm) with respect to the parallel (||) and perpendicular ($\perp$) components of the nematic director $\vec{n}$, respectively. A Glan prism polarizer was placed parallel (*x*-axis) or perpendicular (*y*-axis) to the incident-side rubbing direction. An AC square-wave with 800 Hz frequency was used to unwind the twisted structure into a homeotropic configuration. Normally-incident transmission spectra were measured with a high-resolution spectrophotometer (Shimadzu UV-3600). A typical appearance of the transmission spectrum considered in this study is presented in Fig. 3(a).

Besides the choice of materials and the cell preparation technology, this experimental setup has some principal distinctions from [19]. First, the electrodes were formed upon the zirconium dioxide layer (between LC and the mirror) to provide a more accurate account of the applied voltage. Second, the defect and the surfactant layers are much thicker. This feature increases the number of defect modes and makes their dynamic behaviors (spectral shift under applied voltage) richer and more complicated.

## IV. COMPARISON OF EXPERIMENTAL AND NUMERICAL RESULTS

The experimental transmission spectrum of PC/TN PBG is shown in Fig. 3(a) for a polarizer oriented transversely to the direction of rubbing on the input mirror. The defect transmittance peaks in the center of PBG are damped by material absorption and light scattering. Four peaks are magnified in Fig. 3(b). In the simulation (dashed line) the following parameters were used: the ITO film ($n_{ITO}$ = 1.88858 + 0.006$i$, $d_{ITO}$ = 140 nm), fused quartz substrate ($n_{Sub}$ = 1.45), alignment layer ($n_{PVA}$ = 1.515, $d_{PVA}$ = 1000 nm), and LC defect layer ($n_e$ = 1.701 + 3.9 × $10^{-4}i$, $n_o$ = 1.536 + 3.9 × $10^{-4}i$, $d$ = 10 815 nm). PVA and LC defect layer thicknesses have been acceptably tuned to match the experimental results. Material aging leads $n_e$ to be adjusted as well. As shown in Fig. 3(b), the agreement between the experimental and the simulated spectrum is exhilarating satisfactory. Although the material absorption has been taken into account, the intensity of defect modes in calculated results is still a little bit higher than the experimental one. The full width at half maximum of simulated defect modes is about 2 nm, which is comparable to the experimental data. The discrepancy between the measured and simulated spectra could attribute to the slight light scattering caused by the structural interface roughness and other experimental uncertainties including imperfect fabrication of the multilayers.

Four transmission peaks in Fig. 3(b) were chosen to illustrate the typical spectral shift under applied voltage (Fig. 4(a)). The shift of transverse components is a characteristic feature of twisted structure. This shift was first reported in [19] and had not been observed for in-plain LC orientation [15-18]. The numbers labeled in Fig. 4(b) is called the mode number of which the concept has been mentioned in the previous works [18]. Each defect-mode peak would have one mode number corresponding to the oscillating number of the electromagnetic waves in the medium; in other words, the number of half waves. In this sense, peaks can be identified by their mode numbers. For example, the left-most peak at 573 nm under no voltage has the same mode number 68 as the peak at 575 nm when $U$ = 10 V. At voltages near the Fréedericksz threshold, the number of peaks is doubled, when the voltages getting higher, the minor peaks conceal into the major peaks. The spectral jump in mode numbers will be discussed in detail later. When the polarizer is parallel to the rubbing direction, longitudinal molecular dipoles are excited. They produce extraordinary refraction which is extremely sensitive to the molecular director and reorientation. Therefore, the longitudinal spectrum is even more complicated and steep (Fig. 5). Owing to the sensitivity of voltage-induced shift, the sampling interval is as small as $\Delta U$ = 0.02 V, which allows the continuity of peak shift to be captured. When the voltage is beyond 1.4 V, a restructuring process occurs. The brighter peaks become less intense and slow down on the voltage-induce shift. The small peaks (or satellites) take over the major place and enlarge the shift degree. Simulation (Fig. 5 (b)) predicts this restructuring at a lower voltage of 1.2 V due to the voltage drop on the surfactant layers is neglected.

## V. INTERPRETATION OF RESULTS

The series of small peak–satellites is observed in the experimental spectrum for a transverse polarizer (Fig. 3(b)). In fact, the wavelengths of these satellites correspond to that of major peaks of longitudinal polarization. This phenomenon can be explained by the experimental uncertainties that were not concerned in the model, such as possible structural domains with different sense of rotation, pretilt angle and the surfactant roughness visible at planar orientation. However, above uncertainties are not the reasons for the existence of satellites, which only enhance the irreducible satellites. Numerical simulation supports this fact qualitatively, but not quantitatively. The amplitude of numerical satellites is several times lower than in the experiment.

Actually, the satellites originate from the elliptic polarization of optical modes generated by the TNC structure. The adiabatic following in TNC leads the linearly-polarized light being rotated along the twisted structure; in other words, the twisted adiabatic mode acts as a polarization waveguide. Considering a rod-like NLC with $n_o < n_e$, within a pair of parallel polarizers, the transmittance $T$ can be expressed as a sinc function of adiabaticity [31,32]

$$T = \left[ \frac{\sin\left(\sqrt{1+u^2}\,\pi/2\right)}{\sqrt{1+u^2}} \right]^2, \tag{11}$$

where $u = 2d\Delta n/\lambda$ is the adiabatic Mauguin parameter, $\Delta n$ is birefringence, $d$ is the thickness of the LC layer, and $\lambda$ is the wavelength of the incident light. The Mauguin condition is $u \gg 1$.

Placing a twisted nematic between the PC mirrors brings about a new physics of the optical resonator. This resonator is conventionally named as defect layer. Resonator modes of this resonator [33] are conventionally named as defect modes. The explanation of spectral shift of transverse mode involves regenerative feedback and phase retardation. These phase retardations of optical modes are not equal to $2\pi d \cdot n_{o,e}/\lambda$. Effective refraction indices differ from ordinary and extraordinary refraction indices of O- and E-modes. To avoid confusion, we introduce a slow semi-longitudinal mode, L-mode, the first type of PC/TN modes. L-mode is similar to E-mode in that its ellipse's long axis is parallel to the LC director. The sense of elliptic rotation of L-mode is opposite to the twist sense of LC director. The other type of PC/TN modes is fast semi-transverse T-mode. The O-mode of non-twisted resonator is similar to T-mode, although O-mode has constant ordinary refraction index and can not be shifted in frequency [12, 15-18]. The sense of elliptic rotation of T-mode is same as the LC director twist sense.

The numerical voltage-dependent transmittance spectrum of an unpolarized light is presented in Fig. 6. Four essential ranges are shown with dashed separators.

I) $0 < U < U_C = 0.78$ V. No dynamics below the Fréedericksz threshold.
II) $0.78$ V $< U < 1.1$ V. The L-mode peak series moves quickly towards the shorter wavelengths. But the T-mode peak series remains virtually at fixed wavelengths due to the Mauguin condition ensuring effective ordinary refraction index (Fig. 1 (a)).
III) $1.1$ V $< U < 2.0$ V. The Mauguin condition breaks down in the thin middle layer (Fig. 1(b)). LC in this stage exhibiting a steep azimuthal rotation and low effective birefringence $\Delta n_{eff} \ll n$. One more thing is that the T-modes interact with L-modes when their parity is opposite, we will discuss about it later.
IV) $2.0$ V $< U$. The voltage now is approximately three times higher than the Fréedericksz threshold. The middle LC layer has a virtually homeotropic director orientation (Fig. 1(c)). The LC layers on the borders are in-plane oriented and there is no twist and no Mauguin rotation of polarization. The L- and T-modes become linearly-polarized. Vertical and horizontal directions without clockwise or counterclockwise twist become symmetric. Peaks of L- and T-modes with the same mode number combine to form doublets. These doublets experience a voltage-induced blueshift as the integral refraction index goes down with LC boundary layers thinning-down.

Let's focus on the stages II and III, from which one can observe the unusual interaction on mode intersections. Figure 7(a) shows the cross of T-mode No. 68 and L-mode No. 74. The difference of numbers is even indicating the same parity. Transmission peaks corresponding to these modes make linear superposition and cross each other under increasing voltage. A rough estimation for the number of half waves inside the LC layer is yielded by $N = 2nd/\lambda$

$$N_T = 2 \cdot 10815/(574/1.536) = 57.9 \approx 68\text{-}10;$$
$$N_L = 2 \cdot 10815/(576/1.701) = 63.9 \approx 74\text{-}10. \qquad (12)$$

We subtract the oscillations in the surfactant layers, 5 oscillations in each layer. Standing wave nodes do not occur exactly at the LC layer borders.

Figure 7(b) illustrates the interaction of T-mode No. 68 with L-mode No. 73, which have the opposite parity. Transmission peaks avoid crossing each other. This crossing-avoid phenomenon is well known in quantum mechanics as level repulsion phenomenon [24]. The crossing is avoided due to mode coupling. In quantum mechanics, the crossing is observed only at the interaction of opposite parity modes [35]. In PC/TN, however, it is contrary: the opposite parity modes avoid crossing because the parity uses the mode spatial symmetry in the direction parallel rather than perpendicular to the transmission channel, and vice versa.

The L-mode refraction index reduces with the increasing voltage, which leads to a blueshift of the L-mode wavelength. In Fig. 7(b), the long-wave peak (L73) shifts towards the shorter one (T68). Albeit the L- and T-mode wavelengths lie within a short distance, both of them are excited at the long-wave peak wavelength with the amplitudes $a_L$ and $a_T$. Their superposition is a mixed mode with the electric field $E_{lw}$ at long-wave peak:

$$E_{lw} = a_L E_L + a_T E_T, \qquad (13)$$

$E_{L,T}$ are field components corresponding to pure L- and T-modes. With the increasing voltage, $a_L$ reduces from unity to zero, while $a_T$ grows from zero to unity. The longitudinal-polarized transmitted light changes to a transverse-polarized light, meanwhile slows down its shift rate. On the contrary, the short-wave peak changes from transversely polarized light to longitudinally polarized light and the peak shift rate increases. More precisely, the interacting modes "exchange" their peaks with each other. By analogy with quantum mechanics, isolated modes are diabatic states and peaks refer to adiabatic superposition states. The number of half-waves inside the defect is not conserved for the transmittance peak. This discrete number varies in a smooth manner due to ellipticity (see in the text further down). The same parity condition is mentioned in [17, 18], but the in-plain orientated PC/NLC situation is different. The in-plain E- and O-modes remain orthogonal while their linearly-polarized projections interfere. Therefore, the in-plain interference requires crossed polarizers. On the opposite, with PC/TN neither orthogonality, nor polarizers are required.

The closer the mode numbers, the stronger the mode interaction with the growing overlap integral, more the distance of level repulsion. The series of avoided crossing regions in Fig. 6 (regions b and d) must be extended to where the voltages higher than 2 V (region f), where the interaction occurs for T-mode and L-mode with the neighboring mode numbers. As a result of this strongest interaction, T-mode No. 68 jumps to the neighboring long-wave peak. L-mode No.

69 jumps to the neighboring short-wave peak. The mode number inside the transmittance peak is shifted to ±1, accordingly.

Figure 8 shows the L-mode local optical field intensity (squared electric field strength averaged in time) inside the PC/TN. Nodes and antinodes of the standing wave are dumped in the middle, which follows from ellipticity of opposite running waves. In the extreme case, two opposite circularly polarized waves of the same sense of rotation interfere without nodes. Nodes of vertical polarization are compensated by antinodes of horizontal polarization, and vise versa. The illustrated oscillation suppression provides a mechanism for a mode to change its integral mode number.

## VI. CONCLUSIONS

We have considered a one-dimensional photonic crystal playing the role of a resonator with a tunable twisted-nematic anisotropic medium. A major experimental fact, the presence of satellite peaks, has been explained owing to the Mauguin condition violation and the interference between L- and T-modes. These series interact with each other without peak crossing in the case of same parity: the mode jumps to the neighbouring transmission peak and changes the mode number of the peak. This one-dimensional PC/TN system presents an extremely simplified example of avoided crossing, which has been well known in the multidimensional, multi-channel systems [35].

Four essential ranges of applied voltages show distinct peculiar physics. No changes are observed when applied voltage is less than the Fréedericksz threshold. The threshold voltage promotes a quick spectral shift of the L-mode series to a shorter wavelength range while the T-mode series remains at fixed wavelengths. At higher voltages the Mauguin condition is broken due to the unwinding effect, which allows interaction of T-modes with L-modes of opposite parity. After the LC twist is removed by a voltage-induced homeotropic layer, the L- and T-modes become linearly polarized and equivalent because of the spatial symmetry. The L- and T-modes of the same number combine to form the doublets. These doublets shift to a homeotropic mode as the integral refraction index goes down to ordinary refraction index.

The PC/TN structure suggest a sort of potential for practical application such as a light valve, monochromatic selector *etc*. Besides of mentioned spectral and electrooptical applications, the high precision method that has been utilized in this study can also provide a new way for obtaining the LC elastic constants through simulation of the inverse problem.

## ACKNOWLEDGMENTS

This work was supported in part by the Grants of DSP No. 2.1.1/3455; RAS Nos. 3.9.1 and 21.1, SB RAS Nos. 5 and 144 and by the National Science Council of Taiwan under Grant No. NSC 98-2923-M-033-001-MY3.

Address correspondence to I. V. Timofeev, L.V. Kirensky Institute of Physics, Krasnoyarsk Scientific Centre, SB RAS, Krasnoyarsk, 660036, Russia. E-mail: tiv@iph.krasn.ru

**Figures**

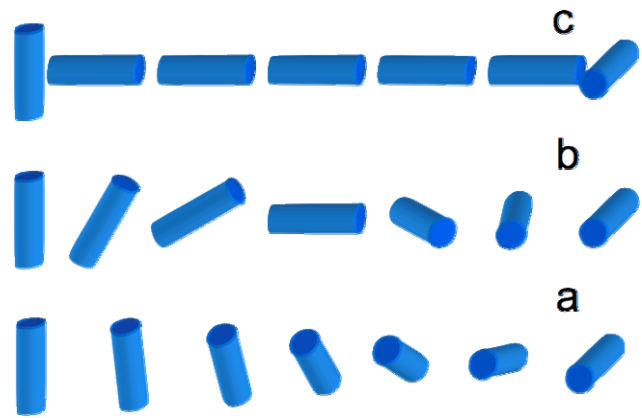

FIG. 1. Schematic of LC reorientation in a TNC under applied voltage. (a) Constant rotation at voltages below the Fréedericksz threshold. (b) Homeotropic lubrication. (c) High voltage reduces boundary planar layers.

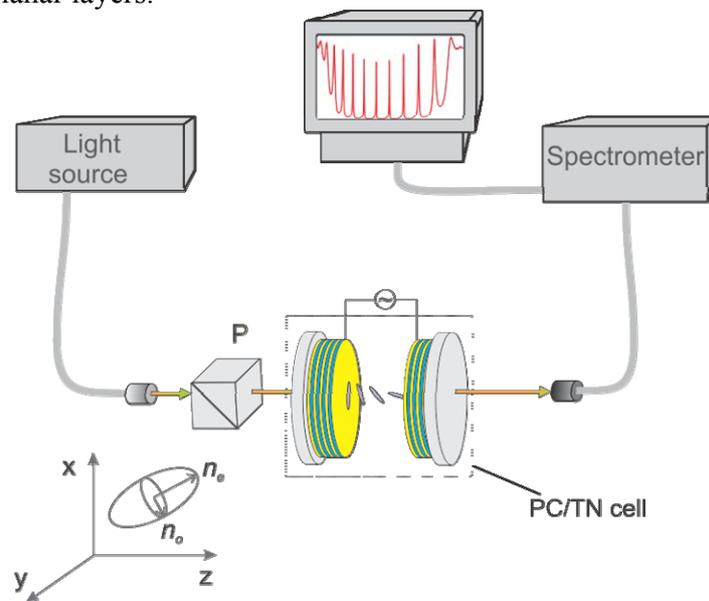

FIG. 2. Experimental layout for the measurement of transmission spectra of a PC/TN cell. The light source and spectrometer blocks are built-in components of the Shimadzu UV-3600 spectrophotometer. **P** is a polarizer

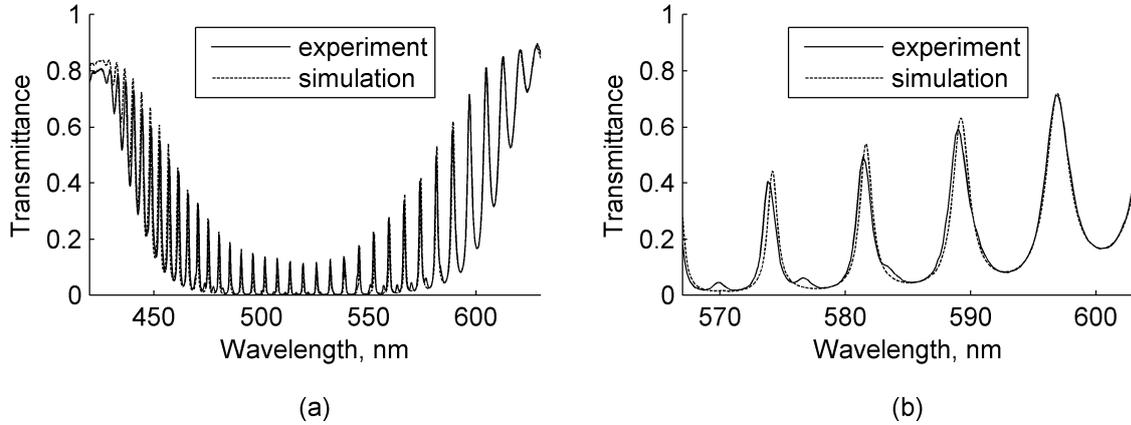

(a) (b)

FIG. 3. Experimental (solid line) and numerical (dashed line) transmission spectra of the PC/TN for transverse polarizer orientation. (a) Spectral data within the PBG. (b) An extended scale showing four typical defect transmittance peaks.

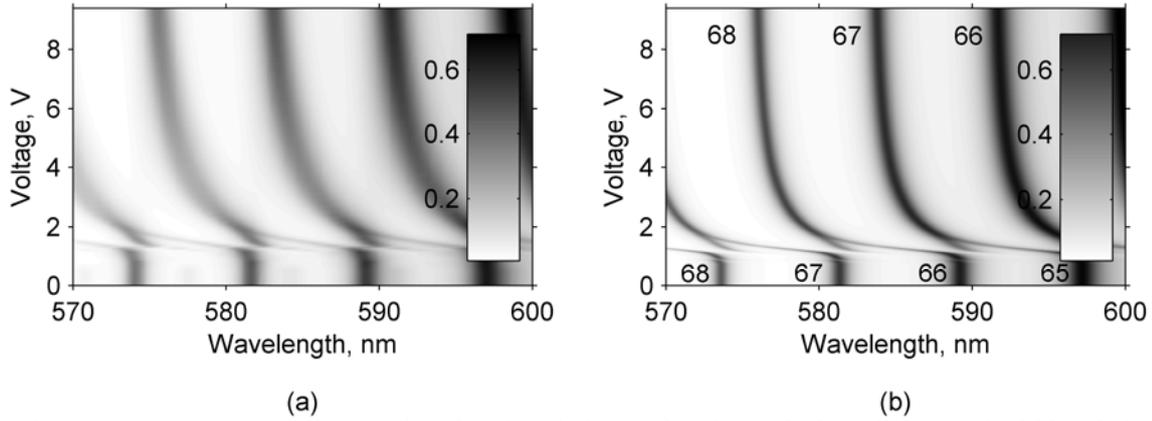

(a) (b)

FIG. 4. (a) Experimental and (b) simulated data of voltage-induced spectral shift of the four transmission peaks as shown in FIG. 3(b). The darker color corresponds to the higher transmittance. The numbers labeled on the top and the bottom in (b) stand for the mode numbers.

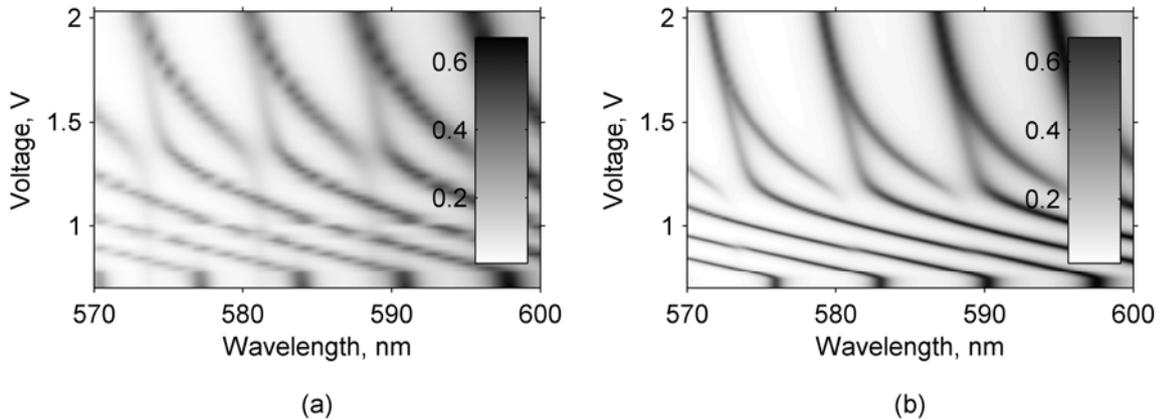

(a) (b)

FIG. 5. Voltage-induced spectral shift of transmittance peaks for longitudinal polarizer orientation mode. Darker color represents higher transmittance. (a) experiment; (b) simulation

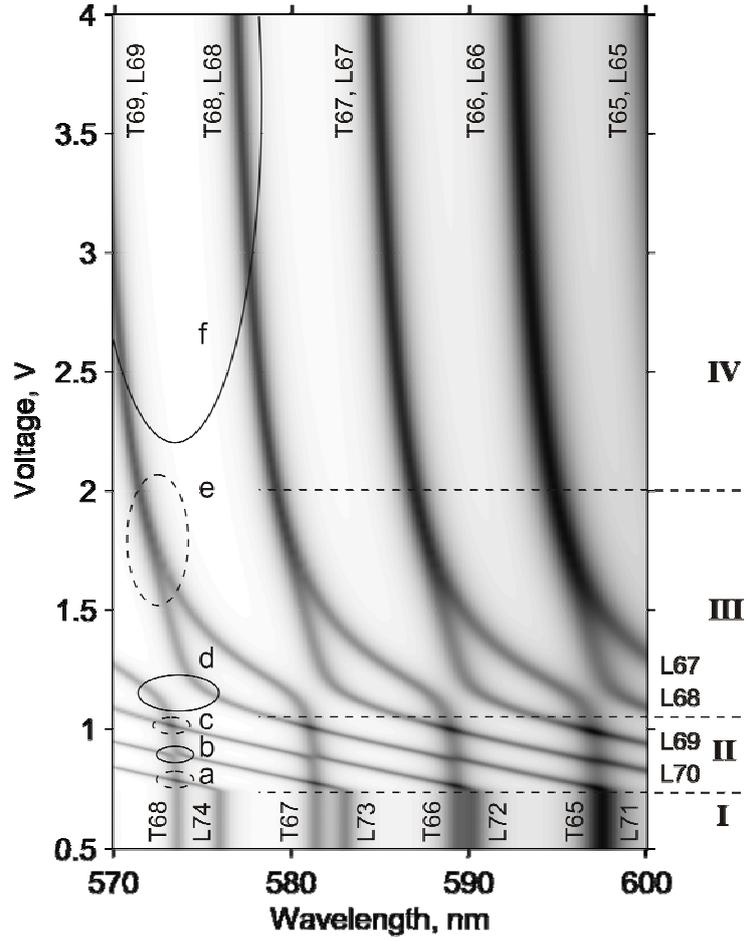

FIG. 6. Calculated transmittance spectrum of unpolarized light versus voltage. Darker color corresponds to higher transmittance. The regions of interaction of T-mode No. 68 with L-modes No. 74 (a), 72 (c), 70 (e) are shown by dashed circles (same mode parity). The regions of interaction of T-mode No. 68 with L-modes No. 73 (b), 71 (d), 69 (f) are shown by solid circles (opposite mode parity). Dashed lines separate four voltage ranges exhibiting different types of mode behavior: below the threshold (I), L-mode shift (II), strong mode interaction (III), and doublets (IV).

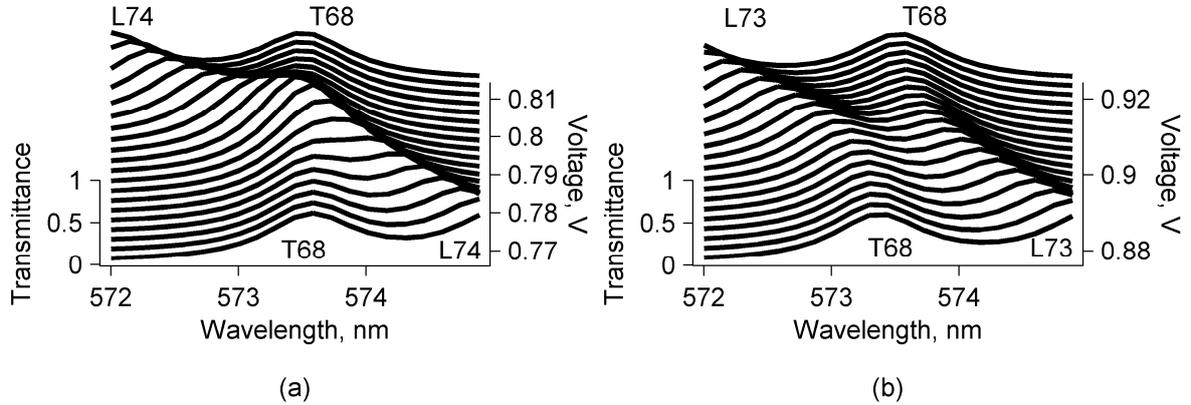

(a) (b)

FIG. 7. The regions of mode interaction as zoomed in from encircled zones in FIG. 6. (a) Same-parity modes cross each other (see FIG. 6, zone a). (b) Opposite-parity modes avoid crossing (see FIG. 6, zone b).

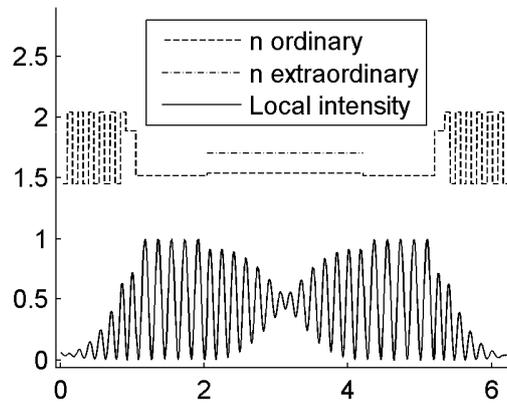

FIG. 8. Spatial distribution of local intensity inside the PC/TN for L-mode at 558.1 nm. Oscillation suppression embarrasses calculation of the mode number. The parameters are the same as used to produce FIG. 3, except for the LC layer of 2.2 $\mu$m thickness, which is 5 times reduced to show the frequent oscillations in detail. The ordinary refraction index is shown by dashed line and the extraordinary refraction index in the NLC layer is indicated by a dash-dot line.